\documentclass{PoS}

\usepackage{epsfig}
\usepackage{amsmath}
\usepackage{amssymb,amsfonts}
\usepackage{graphicx}
\usepackage{graphics}
\usepackage{amsthm}
\usepackage{graphicx}
\usepackage{multirow}
\usepackage{subfigure}
\usepackage{float}

\newcommand{\cA}{{\cal A}}
\newcommand{\cAb}{{\overline{\cal A}}}
\newcommand{\cF}{{\cal F}}
\newcommand{\cFb}{{\overline{\cal F}}}
\newcommand{\cD}{{\cal D}}
\newcommand{\cDb}{{\overline{\cal D}}}
\newcommand{\cQ}{{\cal Q}}
\newcommand{\cU}{{\cal U}}
\newcommand{\cN}{{\cal N}}
\newcommand{\cUb}{{\overline{\cal U}}}
\newcommand{\Tr}{{\rm Tr\;}}

\newcommand{\vn}{ {\bf n} }

\newcommand{\hatbmu}{\widehat{\boldsymbol {\mu}}}
\newcommand{\hatbe}{\widehat{\boldsymbol {e}}}

\def\bec{\begin{center}}
\def\eec{\end{center}}
\def\beq{\begin{equation}}
\def\eeq{\end{equation}}
\def\bea{\begin{eqnarray}}
\def\eea{\end{eqnarray}}

\title{Investigating the sign problem for two-dimensional $\mathcal{N}=(2,2)$ and $\mathcal{N}=(8,8)$ lattice super Yang--Mills theories}

\ShortTitle{On the sign problem for 2D $\mathcal{N}=(2,2)$ and $\mathcal{N}=(8,8)$ lattice SYM theories}

\author{\speaker{Richard Galvez}\thanks{This work is supported in part by the U.S. Department of Energy grant under contract no. DE-FG02-85ER40237 and Science Foundation Ireland grant 08/RFP/PHY1462. Simulations were performed using USQCD resources at Fermilab. AJ's work is also supported in part by the LDRD program at the Los Alamos National Laboratory.}\\
        Department of Physics, Syracuse University, Syracuse, NY 13244, USA.\\
        E-mail: \email{ragalvez@syr.edu}}

\author{Simon Catterall\\
        Department of Physics, Syracuse University, Syracuse, NY 13244, USA.\\
        E-mail: \email{smc@physics.syr.edu}}

\author{Anosh Joseph\\
        Theoretical Division, Los Alamos National Laboratory, Los Alamos, New Mexico 87545, USA\\
        E-mail: \email{anosh@lanl.gov}}

\author{Dhagash Mehta\\
        Department of Physics, Syracuse University, Syracuse, NY 13244, USA.\\
        E-mail: \email{dbmehta@syr.edu}}

\abstract{Recently there has been some controversy in the literature concerning the existence of a fermion sign problem in the $\cN= (2,2)$ supersymmetric Yang--Mills (SYM) theories on the lattice. In this work, we address this issue by conducting Monte Carlo simulations not only for $\cN=(2, 2)$ but also for $\cN=(8, 8)$ SYM in two dimensions for the $U(N)$ theories with $N=2$, using the new ideas derived from topological twisting followed by geometric discretization. Our results from simulations provide the evidence that these theories do {\it not} suffer from a sign problem as the continuum limit is approached. These results thus boost confidence that these new lattice formulations can be used successfully to explore the nonperturbative aspects of the four-dimensional $\cN=4$ SYM theory.}

\FullConference{ The XXIX International Symposium on Lattice Field Theory - Lattice 2011\\
July 10-16, 2011\\
Lake Tahoe, California, USA}

\begin{document}
%-----------------------------
\section{Introduction}
\label{sec:intro}
\vspace{-10 pt}
%-----------------------------
Supersymmetric Yang--Mills (SYM) theories form an interesting class of field theories from a variety of perspectives. They play a role  as toy models for understanding the properties of nonabelian gauge theories including QCD. They could be strong candidates for describing the physics beyond the Standard Model. Some of these gauge theories also play an important role in the AdS/CFT correspondence, and thus are connected to a class of gravitational theories. Many interesting features are exhibited by these theories, for example, dynamical supersymmetry breaking, are inherently non-perturbative in nature. To study such non-perturbative features we are motivated to regularize them on a space-time lattice.

Unfortunately, conventional regularization methods for supersymmetric field theories on the lattice have been proven difficult or even impossible in most cases. This difficulty can be traced back to the fact that the supersymmetry algebra, which is an extension of the usual Poincar\'e algebra, is broken completely by naive discretization on a space-time lattice. Thus, one has to search for nontrivial methods to discretize these theories which respect at least part of the supersymmetry algebra. It turns out that Nature has provided us a set of elegant tools, topological twisting and orbifolding, to construct at least certain classes of supersymmetric theories on the lattice while preserving a subset of the continuum supersymmetry algebra. The reviews \cite{Reviews} and references therein provide details of these techniques, which, we will be using in this work. There also exist other complementary approaches in the literature and they can be found in \cite{Sugino:2003yb-Sugino:2004qd-hep-lat/0507029-arXiv:0707.3533-Kanamori:2008bk-Hanada:2009hq-Hanada:2010kt-Hanada:2010gs-Hanada:2011qx}.

Interestingly, several continuum SYM theories, including the well known $\cN=4$ SYM in four dimensions, can be implemented on a space-time lattice via geometric discretization of the corresponding topologically twisted forms of these theories. On a flat Euclidean space-time, these classes of continuum SYM theories and their topologically twisted cousins are related just through a change of field variables.

Lattice theories constructed this way using these techniques have no doubling problem, respect gauge-invariance, preserve a subset of the original supersymmetries and target the usual continuum theories in the naive continuum limit. However, only certain classes of lattice SYM theories are possible with this construction scheme: the requirement is that the target SYM theories must possess a sufficient number of extended supersymmetries. To be more precise, the number of supercharges must be an integer multiple of $2^D$ where $D$ is the space-time dimension. This includes the $\cN=(2, 2)$ SYM theory in two dimensions and $\cN = 4$ SYM in four dimensions. In this work, we study these two theories, reduced in the case of the $\cN=4$ model to two dimensions, yielding the $\cN=(8,8)$ SYM theory.

Even with the existence of a supersymmetric lattice construction for a given SYM theory, one might encounter another difficulty that would prevent one from extracting sensible results from lattice simulations. This difficulty is known as the fermionic sign problem. Consider a generic lattice field theory with a set of bosonic and fermionic degrees of freedom, $\phi$ and $\psi$ respectively. Then the partition function of the theory can be written as $Z = \int [D\phi][D\psi]~\exp (-S_B[\phi] - \psi^T M[\phi] \psi)$. After integrating out the fermionic degrees of freedom, we have $Z = \int [D\phi]~{\rm Pf}(M)~\exp (-S_B[\phi])$. The matrix $M$ corresponds to the fermion operator and ${\rm Pf}(M)$ the corresponding Pfaffian. For a $2n \times 2n$ matrix $M$, the Pfaffian is explicitly given as ${\rm Pf}(M)^{2} = \mbox{Det}\, M$. In the supersymmetric lattice constructions which we will consider in this work the matrix $M$ at non zero lattice spacing is a complex operator, and one might worry that the resulting Pfaffian could exhibit a fluctuating phase depending on the background boson fields $\phi$. Since the Monte Carlo simulations must be performed with a positive definite measure, the only way to incorporate this fluctuating phase is through a reweighting procedure, which folds this phase in with the observables of the theory. Thus the expectation values of the observables derived from such simulations can be contaminated by drastic statistical errors, which could overwhelm the values of the quantities we are trying to measure.

In the constructions of supersymmetric lattice gauge theories, there has been an ongoing debate on the existence of a sign problem in the two-dimensional $\cN=(2,2)$ lattice SYM \cite{Giedt:2003ve, Catterall:2008dv, Hanada:2010qg}. In \cite{Giedt:2003ve}, it was shown that there is a potential sign problem in the two-dimensional $\cN = (2, 2)$ lattice SYM. Furthermore, in \cite{Catterall:2008dv} numerical evidence was presented of a sign problem in a phase quenched dynamical simulation of the theory at non-zero lattice spacing. More recently Hanada {\it et al.} \cite{Hanada:2010qg} have argued that there is no sign problem for this lattice theory as the continuum limit is approached. However, the models studied by these various groups differed in detail; Catterall {\it et al.} studied an $SU(2)$ model obtained by truncating the supersymmetric $U(2)$ theory and utilized bosonic link fields valued in the group $SL(2,C)$, while Hanada {\it et al.} used a $U(2)$ model where the complexified bosonic variables take their values in the algebra of $U(2)$ together with the inclusion of supplementary mass terms to control the fluctuations of the scalar fields in the theory.

In this work, we present results from simulations of the two dimensional $\cN=(2, 2)$ $U(2)$ SYM theory (which we will refer to from now on as the $\cQ = 4$ theory, with $\cQ$ the number of supercharges) and the maximally supersymmetric $\cN = (8, 8)$ $U(N)$ SYM theory (we refer to this theory as the $\cQ = 16$ theory) using algebra based parameterizations of the bosonic link variables. We measure the average phase fluctuations as we approach the continuum limit. In all cases we find evidence that the phase fluctuations disappear in the continuum limit indicating
the absence of the sign problem in these lattice theories as the continuum limit is approached.
%-----------------------------------------------------------
\section{Supersymmetric Yang--Mills theories on the lattice}
\label{sec:SYM-on-lattice}
\vspace{-10 pt}
%-----------------------------------------------------------
As mentioned before we will be studying the lattice SYM theories constructed using the twisted approach followed by geometric discretization. At this point we also stress that the lattice SYM theories constructed using orbifold methods are equivalent to twisted constructions \cite{Unsal:2006qp-Catterall:2007kn-Damgaard:2007xi}. The idea of twisting \cite{Witten:1988ze} is to decompose the fields of the Euclidean SYM theory in $D$ spacetime dimensions in representations not in terms of the original (Euclidean) rotational symmetry $SO_{\rm rot}(D)$, but a twisted rotational symmetry, which is the diagonal subgroup of this symmetry and an $SO_{\rm R}(D)$ subgroup of the R-symmetry of the theory, that is, $SO(D)^\prime={\rm diag}(SO_{\rm Lorentz}(D)\times SO_{\rm R}(D))$. It should be noted that the R-symmetry group of the theory must be large enough to contain $SO_{\rm R}(D)$ as a subgroup. Otherwise, the twisted lattice constructions would not turn out to be successful.
%------------------------------------------------------
%\subsection{Two-dimensional $\cQ=4$ lattice SYM theory}
%\label{sec:2d-formulation}
%------------------------------------------------------

{\bf Two-dimensional $\cQ=4$ lattice SYM theory}: The two-dimensional $\cQ = 4$ SYM theory is the simplest example of a gauge theory that permits topological twisting and thus satisfies our requirements for a supersymmetric lattice construction. Its R-symmetry possesses an $SO(2)$ subgroup corresponding to rotations of its two degenerate Majorana fermions into each other. After twisting, we end up with fermionic fields ($\eta$, $\psi_a$, $\chi_{ab}$) and a set of complexified bosonic fields ($\cA_a$, $\cAb_a$). The prescription for discretization is somewhat natural. The complexified bosonic fields are represented as complexified Wilson gauge links $\cA_a(x) \rightarrow \cU_a(\vn)$, living on the links of a lattice, which for the moment can be thought of as hypercubic, with integer-valued basis vectors $\hatbmu_1 = (1, 0),~~~\hatbmu_2 = (0, 1)~$. They transform in the usual way under $U(N)$ lattice gauge transformations $\cU_a(\vn)\to G(\vn)\cU_a(\vn)G^\dagger(\vn)~$. Supersymmetric invariance then implies that $\psi_a(\vn)$ live on the same links and transform identically. The scalar fermion $\eta(\vn)$ is clearly most naturally associated with a site and transforms accordingly $\eta(\vn)\to G(\vn)\eta(\vn)G^\dagger(\vn)~$. The two-form field $\chi_{ab}$ lives on the diagonal link. The field orientations are chosen such that gauge invariance is preserved on the lattice.

The continuum derivatives are replaced by difference operators. Motivated by the natural technology developed for applying the derivative operators to arbitrary lattice p-forms \cite{Aratyn:1984bd}, we need just two derivatives given by the expressions:
\bea
\cD^{(+)}_a f_b(\vn) &=& \cU_a(\vn)f_b(\vn + \hatbmu_a) - f_b(\vn)\cU_a(\vn+ \hatbmu_b)~,\\
\cDb^{(-)}_a f_a(\vn) &=& f_a(\vn)\cUb_a(\vn)-\cUb_a(\vn - \hatbmu_a)f_a(\vn - \hatbmu_a)~.
\eea
The lattice field strength is then given by the gauged forward difference acting on the link field: $\cF_{ab}(\vn) = \cD^{(+)}_a \cU_b(\vn)$, and is automatically antisymmetric in its indices. Similarly the covariant backward difference appearing in $\cDb^{(-)}_a \cU_a(\vn)$ transforms as a 0-form or site field and hence can be contracted with the site field $\eta(\vn)$ to yield a gauge invariant expression.

The final lattice version of the action is
\beq
\label{eq:2d-latt-action}
S = \sum_{\vn} \Tr \Big(\cF_{ab}^{\dagger}(\vn) \cF_{ab}(\vn) + \frac{1}{2}\Big(\cDb_a^{(-)}\cU_a(\vn)\Big)^2 - \chi_{ab}(\vn) \cD^{(+)}_{[a}\psi_{b]}(\vn) - \eta(\vn) \cDb^{(-)}_a\psi_a(\vn) \Big)~,
\eeq
which can be shown to be $\cQ$-exact same as its continuum counterpart.
%------------------------------------------------------
%\subsection{Four-dimensional $\cQ=16$ lattice SYM theory}
%\label{sec:4d-lattice-theory}
%------------------------------------------------------

{\bf Four-dimensional $\cQ=16$ lattice SYM theory}: The lattice action of this theory contains a $\cQ$-exact term of precisely the same form as the two-dimensional theory provided in Eq. (\ref{eq:2d-latt-action}) if one extends the indices labeling the fields to run now from one to five. In addition, the appropriate twist (called the Marcus twist \cite{Marcus:1995mq}) of $\cN=4$ YM requires a new $\cQ$-closed term, which was not possible in the two-dimensional theory, whose lattice version is:
\beq
S_{\rm closed} = -\frac{1}{8}\sum_{\vn} \Tr \epsilon_{mnpqr} \chi_{qr}(\vn + \hatbmu_m + \hatbmu_n + \hatbmu_p)
\cDb^{(-)}_p\chi_{mn}(\vn + \hatbmu_p)~,
\eeq
and can be seen to be supersymmetric since the lattice field strength satisfies an exact Bianchi identity \cite{Aratyn:1984bd}: $\epsilon_{mnpqr}\cDb^{(+)}_p\cFb_{qr} = 0$.

In the case of the $\cN=4$ theory, the resulting lattice has a nontrivial structure: It is known as the $A_4^*$-lattice. The lattice is constructed from the set of five basis vectors $\hatbe_a$ pointing out from the center of a four-dimensional equilateral simplex out to its vertices together with their inverses $ -\hatbe_a$. Complexified Wilson gauge link variables $\cU_a$ are placed on these links together with their $\cQ$-superpartners $\psi_a$. Another 10 fermions are associated with the diagonal links $\hatbe_a + \hatbe_b$ with $a>b$. Finally, the exact scalar supersymmetry implies the existence of a single fermion for every lattice site. It is invariant under the exact scalar supersymmetry $\cQ$, lattice gauge transformations $G(\vn)$, and a global permutation symmetry $S^5$, and can be proven free of fermion doubling problems. The $\cQ$-exact part of the lattice action is again given by Eq. (\ref{eq:2d-latt-action}) where the indices $a, b$ now correspond to the indices labeling the five basis vectors of the $A_4^*$ lattice.

The renormalization of this theory has been recently studied in perturbation theory with some remarkable conclusions \cite{Catterall:2011pd}.
%------------------------------------------------------------------
%\subsection{Gauge link parameterizations}
%\label{sec:gauge-link-params}
%------------------------------------------------------------------

{\bf Gauge link parameterizations}: There exist two distinct parameterizations of the gauge fields on the lattice and they both have been used in various simulations of lattice SYM theories. The first one follows the standard Wilson prescription where the complexified gauge fields in the continuum are mapped to link fields $\cU_a(\vn)$ living on the link between $\vn$ and in $\vn + \hatbmu_a$ through the mapping $
\cU_a(\vn) = e^{\cA_a(\vn)}$, where $\cA_a (\vn) = \sum_{i=1}^{N_G} \cA_a^i T^i$ and $T^i=1, \ldots, N_G$ are the anti-hermitian generators of $U(N)$. The resultant gauge links belong to $GL(N,C)$. We call this realization of the bosonic links the {\it exponential or group based parametrization}\footnote{Notice that our lattice gauge fields are dimensionless and hence contain an implicit factor of the lattice spacing $a$.}.

The other parametrization of the bosonic link fields that has been used, particularly in the orbifold literature, simply takes the complex gauge links as taking values in the algebra of the $U(N)$ group\footnote{In fact, a non-compact parametrization of the gauge-fields is also recently used to restore the BRST symmetry on the lattice in Ref. \cite{arXiv:0710.2410}, i.e., to evade the so-called Neuberger $0/0$ problem \cite{Print-86-0394} (see also Refs.~\cite{arXiv:0710.2410} and \cite{arXiv:0912.0450} for the recent progress).}. To obtain the correct lattice kinetic terms needed to target the continuum theory one must then expand the fields around a particular point in the moduli space of the theory corresponding to giving an expectation value to the trace mode of the imaginary part of the link field -- a field which can be identified as a $U(1)$ scalar field in the untwisted theory. The expansion is: $
\cU_a(\vn) = {\mathcal I}+\cA_a(\vn)$. Usually the use of such an algebra based or {\it non compact} parametrization would signal a breaking of lattice gauge invariance. It is only possible here because the bosonic fields take values in a complexified $U(N)$ theory -- so that the unit matrix ${\mathcal I}$ appearing in the expansion above can be interpreted as the expectation value of a {\it dynamical field}. We will refer to this parametrization as the {\it linear or algebra based parametrization}.

Both parameterizations of the gauge links are equivalent at leading order in the lattice spacing, yield the same lattice action and can be considered as providing equally valid representations of the lattice theory at the classical level. In this work we have concentrated on the linear parametrization principally because it is naturally associated with a manifestly
supersymmetric measure in the path integral - the flat measure.
%--------------------------------------
%\section{The correct continuum limit}
%\label{sec:correct-continuum-limit}
%--------------------------------------

{\bf The correct continuum limit}: The requirement that the theory target the correct continuum theory requires that the fluctuations of all dimensionless lattice fields should vanish in the continuum limit. In addition the linear parameterization only yields the correct naive continuum limit if the trace mode of the scalars develops a vacuum expectation value so that appropriate kinetic terms are generated in the classical action. Since no classical scalar potential is present in the lattice theory it is crucial to add {\it by hand} a suitable gauge invariant potential to ensure this feature.
\begin{figure}[t]
\bec \vspace{-0.75cm}
\includegraphics[width=7.5cm, height=4.4cm]{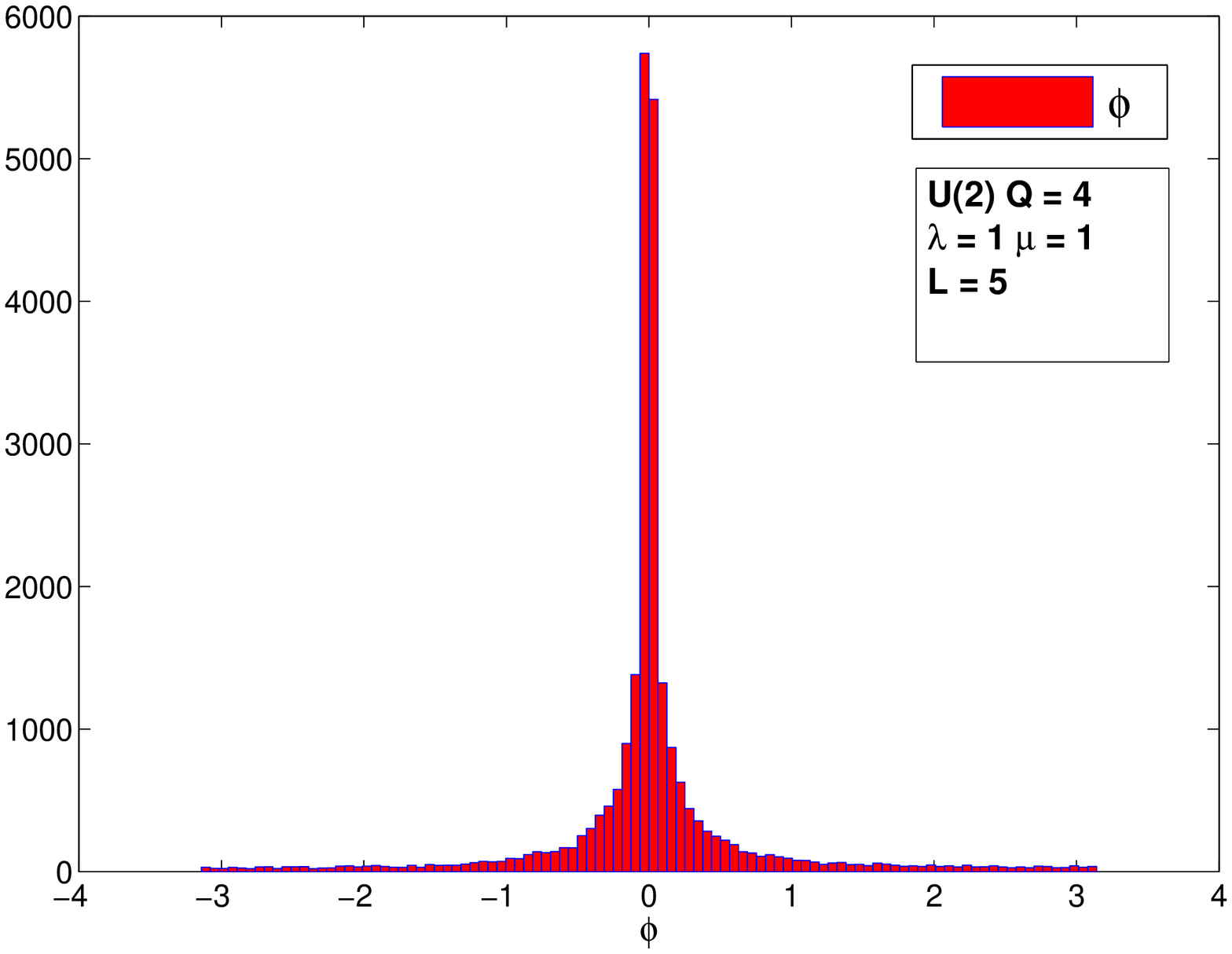}~~\includegraphics[width=7.5cm, height=4.4cm]{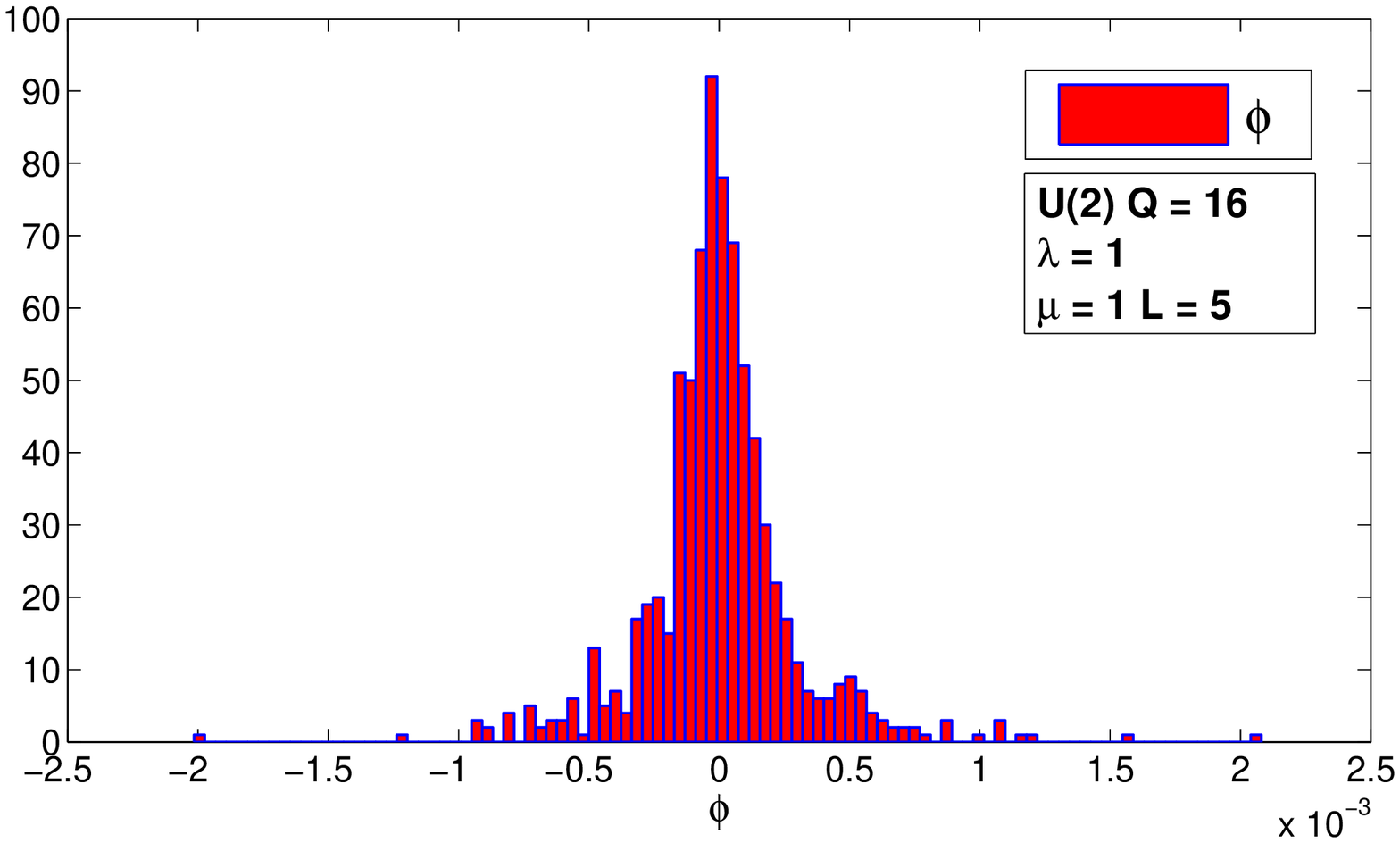}
\vspace{-0.78cm}
\caption{\label{fig:U2_Q4_Q16_2D_hist}The histograms of the phase angles $\alpha$ for $\cQ =4, 16$ theories with gauge group $U(2)$, $\lambda = 1$, $L=5$.}
\eec
\vspace{-32 pt}
\end{figure}
\begin{figure}[t]
\bec
\vspace{-0.4cm} \includegraphics[scale=0.34]{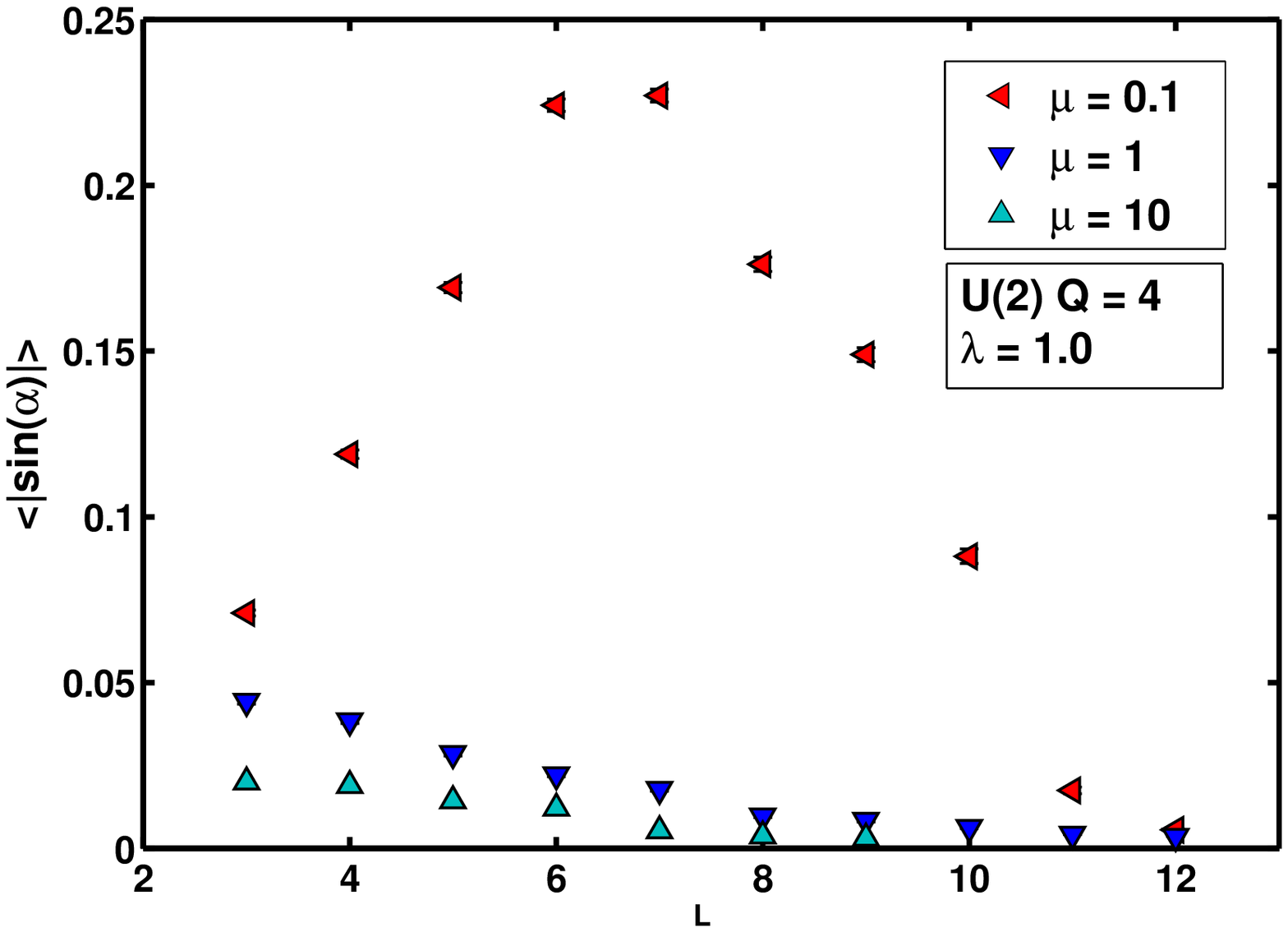}~~~~~~~~~\includegraphics[scale=0.4]{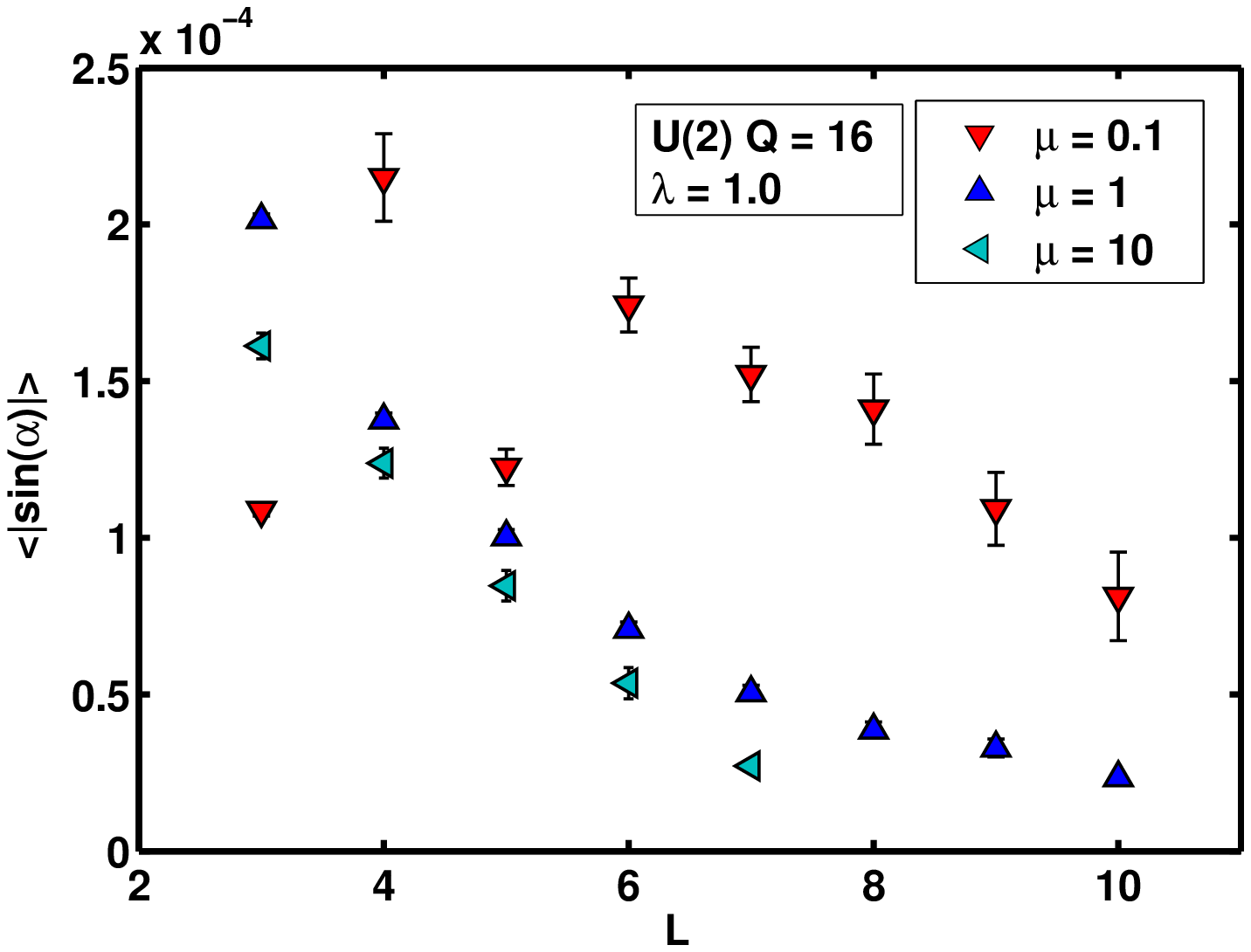}
\vspace{-0.4cm}
\caption{\label{fig:U2_Q4_Q16_2D}The Pfafian phase $<|\sin{\alpha}|>$ for $\cQ =4, 16$ theories with gauge group $U(2)$ and $\mu = 0.1, 1, 10$.}
\eec \vspace{-0.99cm}
\end{figure}

Specifically we add a potential term of the form $S_M = \mu^2 \sum_x \left(\frac{1}{N}{\rm Tr}(\cU_a^\dagger(x) \cU_a(x))-1\right)^2~$ to the lattice action \cite{Hanada:2010qg}. Here $\mu$ is a tunable mass parameter, which can be used to control the expectation values and fluctuations of the lattice fields. Notice that such a potential
obviously breaks supersymmetry -- however all supersymmetry breaking counter terms induced via quantum effects will possess couplings that vanish as $\mu \to 0$ and so can be removed by sending $\mu \to 0$ at the end of the calculation.

In our simulations, we have rescaled all lattice fields by powers of the lattice spacing to make them dimensionless. This leads to an overall dimensionless coupling parameter of the form $N/(2\lambda a^2)$, where $a = \beta/T$ is the lattice spacing, $\beta$ is the physical extent of the lattice in the Euclidean time direction and $T$ is the number of lattice sites in the time-direction. Thus, the lattice coupling is $\kappa = \frac{NT^2}{2t}~$, with $t=\lambda\beta^2$, for the symmetric two-dimensional lattice where the spatial length $L = T$. Note that $t$ is the dimensionless physical `t Hooft coupling in units of the area. In our simulations, the continuum limit can be approached by fixing $t$ and $N$ and increasing the number of lattice points $L\rightarrow\infty$. We have taken three different values for this coupling  $t = 0.5, 1.0, 2.0$ and lattice sizes ranging from $L = 2, \cdots, 12$. We fix the value of $\beta = 1$ so that $t = \lambda$. Theories with $U(N)$ gauge group with $N = 2$ have been examined.

The simulations are performed using anti-periodic (thermal) boundary conditions for the fermions (This forbids exact zero modes that are otherwise present in the fermionic sector). An RHMC algorithm was used for the simulations as described in \cite{Catterall:2011ce}.

In Fig.~(\ref{fig:U2_Q4_Q16_2D_hist}), we show histograms of the phase angles $\alpha$ for different $L$ and $\mu = 0.1,1,10$ for both  $\cQ =4$ and $\cQ= 16$ theories (for $\lambda=1$ and $0.5$, respectively)  with gauge group $U(2)$. In Fig.~(\ref{fig:U2_Q4_Q16_2D}) we show the results for the absolute value of the (sine of) the Pfaffian phase\footnote{Obviously, the absolute value of sine of the angle does not distinguish between, e.g., $0$ and $\pi$. However, focusing on a single measure of the magnitude of the phase fluctuations does allow for a clean extrapolation of to the continuum limit and, as our previous histograms show, suffers from no ambuiguity.}. Three values of $\mu$ are shown corresponding to $\mu = 0.1, 1, 10$. While modest phase fluctuations are seen for small lattices for the case of $\mu = 0.1$ we see that they disappear as the continuum limit is taken.
%--------------------
\section{Conclusions}
\label{sec:conclusions}
\vspace{-10 pt}
%--------------------
We have performed numerical simulations of the four and sixteen supercharge lattice SYM theories in two dimensions to investigate the occurrence of a sign problem in these theories. In contrast to the usual situation in lattice gauge theory, we utilize a non compact parameterization of the gauge fields in which the lattice fields are expanded on the algebra of the group. We have examined both supersymmetric theories for several values of the dimensionless 't Hooft coupling $\lambda \beta^2$ and for the gauge group $U(2)$. We take a careful continuum limit by simulating the theories over a range of lattice size $L = 2 - 12$. In both cases we observe that the average Pfaffian phase is small and goes to zero for a fixed gauge invariant potential as the continuum limit is taken. Indeed, in practice it is sufficiently small even on coarse lattices that there is no need to use a reweighting procedure to compute expectation values of observables. The detailed analysis and results (including for the $U(4)$ theories) are provided in \cite{Catterall:2011aa}.

\end{document}